# Solitons and spin transport in an antiferromagnetic spin chain

Nan-Hong Kuo[1] Sujit Sarkar[2] and C. D. Hu[1,3]

[1]Department of Physics, National Taiwan University, Taipei, Taiwan, R.O.C.

[2]PoornaPrajna Institute of Scientific Research, 4 Sadashivanagar, Bangalore-5600 80, India

[3]Center for Theoretical Sciences, National Taiwan University, Taipei,Taiwan, R.O.C.

Abstract

We study the spin transport on a S=1/2 antiferromagnetic chain with external fields which provids a phase angle. The equation of motion becomes the sine-Gordon equation after Jordan-Wigner transformation and bosonization. Soliton solutions of the sine-Gordon equations for a system of finite length are found and the quantum fluctuations are calculated. The spin is transported as the soliton solutions evolves with the adiabatical variation of the phase angle in the phase space. We observe that the spin is transported by $\Delta S = 1$, as the phase angle changes by the period of $2\pi$ when the quantum fluctuation enables the system makes transition between two solutions. This quantum spin transport is not affected by disturbance of environment.

# 1. Introduction

One-dimensional systems has been studied for a long time [1]. For many physical systems, the equation of motion can be transformed into the sine-Gordon equation. If so, it is very much advantageous because the solutions [2] are well-known and thus, the physical properties can be understood very clearly. However, it is less investigated if the systems have finite length and certain boundary conditions. It can be shown that the solutions are still solitons [3]. If the original sysem is a spin chain, then the solitons are related to some spin states. A natural question would be how the spins change or move with solitons. In most cases, spin transport [4] means that spins are carried by mobil electrons and transported across the samples. However, is this work we present the spin transport in insulators. We consider a Heisenberg spin chain. There is *no* mobile electrons. The spins are transported by the exchange interaction between localized spins with the help of external field.

For a one-dimensional system, an external field can be introduced such that the system can vary in an enhanced parameter space in an adiabatic processes. It was suggested by Shindo [5] that the combination of a dimer field and a magnetic field can give rise to a phase parameter $\varphi$. It can be varied adiabatically to transport spins. Usually, an adiabatic process should be slow enough so as not to induce transition between states. In the spin system, the time scale of the adiabatic process $T$ should be greater than $\hbar/E_g$ where $E_g$ is the spin-gap. As a result, spins rotate slowly with the external field (adiabatic parameter) and spin-flip processes are not considered. However, in certain cases, very interesting phenomena can be found when two bands cross each other. This crossing occurs when at a certain stage of adiabatic variation, the Hamiltonian acquires time-reversal symmetry, and hence, Kramers degeneracy.

Our work was inspired by the seminal work of the quantum particle transport of Thouless [6]. In the case of mobile electrons there are many exotic physical phenomena, including quantum Hall effect, quantum spin Hall effect and topological insulators [7-9]. In these cases, the amount of physical quantity transported is quantized. The inspiration came also from the work of Hasugai on quantum Hall effect [10,11] where the edge states were shown to play an important role. Shindou [5] studied an one dimensional spin chain and concluded that the origin of spin transport is due to the edge state of the system. Kane and Mele [8] and Fu and Kane [12] studied the similar problem. Among their contribution, they found that whether the edge states cross each other is essential for spin transport and the entire process has $Z_2$ symmetry.

Hence, we have two questions at hand. First, whether there is quantized spin transport. The closed line integral of $\mathbf{A} = \nabla\varphi$ yields an integer due to the singularity at the origin. This is exactly the quantization of particle transport proposed by Thouless and Niu and Thouless [6,7]. However, the quantized spin transport is more complicated than quantized charge transport in that the spin polarization need not be an integer or half integer. Intrinsic interaction or perturbation from environment can easily destroy the quantization of spin transport. Therefore, in order to have robust quantum spin transport, the system must possess certain special property. However, as we shall see, quantized pin transport can occur in our system. The second question is how the spin is transported, by the edge state or bulk state? Our conclusion is that the bulk states play an indispensable role. They not only transport spins but are integral part of a topological circle.

Our previous work [13] is an analysis of the non-interacting case. We reached conclusion that spin transport is nothing but the motion of solitons in the system, and hence, through bulk states. In this work, we extended the scope of study to interacting systems. What we have found is that spin transport is through both bulk state and edge states and it is related closely to the topology of solitons and thus render quantum spin transport viable. We also explain the spin transport crossover from bulk state to edge state through the analysis of different solutions of the equation of motion. Equally important is that the quantum characteristic of the spin transport is due to the topological properties of the solution.

We present our work as the following. In section 2, we start with the Heisenberg Hamiltonian of a spin chain. The equation of motion is reduced to the sine-Gordon equation. In section 3 we discuss the boundary condition of a spin chain with finite length. The solutions are given. They are actually solitons and can be expressed in terms of Jacobi elliptic functions. The spectrum of solitons are presented in section 4. It has been found that there is a level-crossing. In section 5 we calculate the quantum fluctuation and their implication. The evolution of the solitons with the adiabatic parameter $\varphi$ is illustrated in section 6. The spin transport is discussed in section 7 and we conclude this work in section 8.

## 2. Equation of motion

The Hamiltonian we considered for a Heisenberg spin (1/2) chain has three parts:

$$H(t) = J\sum_{i=1}^{N} \mathbf{S}_i \cdot \mathbf{S}_{i+1} + h_{st}(t)\sum_{i=1}^{N}(-1)^i S_i^z \qquad 1$$
$$+ \frac{\Delta(t)}{2}\sum_{i=1}^{N}(-1)^i(S_i^+ S_{i+1}^- + S_i^- S_{i+1}^+)$$

where $h_{st}(t)$ is a staggered external magnetic field and the last term is an alternating dimer bonding term which can be induced by applying an electric field to the spin chain to alter the exchange interaction [5]. The time-dependent bond strength $\Delta(t)$ and staggered field $h_{st}(t)$ can be varied adiabatically so as to create a parameter space. We write $\Delta$ and $h_{st}$ as $(h_{st}, \Delta) = R(\cos\varphi, \sin\varphi)$, with $R$ fixed. By varying $\varphi$ adiabatically, we expect spins to be transported.

The method of bosonization has been used successfully to treat various one-dimensional systems [1] including the spin chains. To apply this method, we first make the Jordan-Winger transformation to represent spins by spinless fermion fields $f_i$ and $f_i^+$. Then, the bosonization of $f_i$ and $f_i^+$ will be performed. The basic aspects of bosonization has been discussed extensively in the literature [1], therefore we only present the bosonized form of the Hamiltonian:

$$H = \int dx\{v[\frac{\pi K}{2}\hat{\Pi}^2 + \frac{1}{2K}(\partial_x\hat{\theta}_+)^2] - \frac{R}{\pi}\sin(\hat{\theta}_+ + \varphi)\}$$

where the lattice constant has been set to be 1, $\hat{\theta}_+$ is bosonization phase and $\hat{\Pi}(x) = (-1/\pi)\partial_x\hat{\theta}_-(x)$ is the conjugate momentum. The velocity is $v = J\sin k_F\sqrt{1 + (4\sin k_F)/\pi}$, and the quantum parameter is $K = 1/\sqrt{1 + (4\sin k_F)/\pi}$, with $K < 1$ denoting repulsive fermions or antiferromagnetic spins. Physical systems are in the range $1 \geq K \geq 1/4$ and this is what we have considered. A term of $\cos 2\hat{\theta}_+$ is irrelevant in the sense of renormalization group analysis and hence, is dropped.

As in the usual bosonization procedure, a new variable is defined: $\theta \equiv \hat{\theta}_+/\sqrt{K}$. With the scaling of space-time $z = \sqrt{R\sqrt{K}/\pi v}\,x$ and $\tau = \sqrt{R\sqrt{K}/v}\,t$, the Hamiltonian has the form:

$$H = \sqrt{\frac{vR\sqrt{K}}{\pi}}\int dz[\frac{1}{2}\Pi^2 + \frac{1}{2}(\partial_z\theta)^2 - \frac{1}{\sqrt{K}}\sin(\sqrt{K}\theta + \varphi)] \qquad 2$$

and the it gives the equation of motion

$$\frac{\partial^2\theta}{\partial\tau^2} - \frac{\partial^2\theta}{\partial z^2} - \cos(\sqrt{K}\theta + \varphi) = 0 \qquad 3$$

if the prefactor of the Hamiltonian is scaled away. Equation (3) is the well-known sine-Gordon equation (SGE). It has been studied in details and has extensive application in many physics fields [2]. Here for our purpose we analyze its solution on a finite length spin chain. The spin transport is manifested by the motion of solitons which are the solutions of SGE. The form of the solutions will depend on the boundary conditions which will be discussed in next section.

## 3. Boundary conditions and solutions

For the analysis of the solutions of different values of $\varphi$, the boundary conditions plays an interesting role. Since $\hat{\theta}_+$ is the original phase, $\hat{\theta}_+(z=0) = 0$ and $\hat{\theta}_+(z=L) = 2\pi$ is a natural choice. They imply that the phase of the fermion is differ by $2\pi$ or the spin operators are identical at both ends. With the scaling $\theta = \hat{\theta}_+/\sqrt{K}$, we should set $\theta(z=0) = 0$ and $\theta(z=L) = 2\pi/\sqrt{K}$. These are the twisted boundary conditions. However, we choose instead the boundary conditions

$$\theta(z = 0) = 0 \quad \text{4a}$$
$$\theta(z = L) = 2\pi \quad \text{4b}$$

They are equivalent to the original ones of $\hat{\theta}_+(z)$. The original ones require that $\hat{\theta}_+(z)$ increases by $2\pi$ as $z$ increases by $L$. On the other hand, eqs. (4) require that $\hat{\theta}_+(z)$ increases by $2\sqrt{K}\pi$. This is equivalent to having a system with shorter length so that $\hat{\theta}_+(z)$ increases by less amount or $\theta(z)$ increases only by $2\pi$ instead of $2\pi/\sqrt{K}$. The physical picture remains the same. We choose the set of the boundary conditions in eqs. (4) because, as we shall see, it is easier to illustrate our result.

SGE is an integrable partial differential equation. Among its solutions we will analyze the static solitons first. The related quantum fluctuation will be discussed later. We can easily find the solution [3] of eq. (3) for a system of finite length $L$

$$\theta(z) = \frac{1}{\sqrt{K}}\{\frac{\pi}{2} - \varphi + 4\tan^{-1}[A_{th}sc(\beta(z-z_0),k)]\}, \quad 5$$

where $sc(u,k)$ is one of the Jacobi elliptic functions (JEF) with modulus $k$, $z_0$ is a reference point of the soliton, $\beta = 1/(1 - A_{th}^2)$ and $|A_{th}| = \sqrt[4]{1-k^2} = \sqrt{k'}$. It is also important to note that $sc(u)$ is a periodic and odd function of $u$ with period $2\mathcal{K}$, with $\mathcal{K}$ being the elliptic integral of the first kind with modulus $k$. The properties of JEF can be found in ref. 14.

The boundary conditions in eqs. (4) requires that the angle $\tan^{-1}[A_{th}sc(\beta z - \beta z_0)]$ increases by $\sqrt{K}\pi/2$ when $z$ increases by $L$. Since the period of tangent function is $\pi$ and that of $sc$ is $2\mathcal{K}$, we found that $\beta L \approx n\sqrt{K}\mathcal{K}$ where $n$ is an integer. Furthermore, the larger the value of $n$, the higher the energy as $\beta$ plays the role of wave vector. It is similar to the case of linear standing waves.

From the boundary conditions, we get $\tan\alpha_1 = A_{th}sc(-\beta z_0, k)$ and $\tan\alpha_2 = A_{th}sc(\beta L - \beta z_0, k)$ where $\alpha_1 = (\varphi - \pi/2)/4$ and $\alpha_2 = (2\sqrt{K}\pi + \varphi - \pi/2)/4$. Here we consider two cases:
case 1, $\tan\alpha_1 \tan\alpha_2 < 0$,
case 2, $\tan\alpha_1 \tan\alpha_2 > 0$.
To facilitate the explanation of the following analysis, the range in case 1 will be called the "allowed region" and that in case 2 the "forbidden region". There is no difficulty in finding solutions in the "allowed region" as they are already given in eq. (5). As for the "forbidden region", the boundary conditions require that the phase $\theta(z)$ increase by $2\pi$ as $z$ increases from 0 to $L$. This cannot be achieved if $sc(-\beta z_0, k)$ and $sc(\beta L - \beta z_0, k)$ have the same sign and there is no node in between. The reason is that for a moderate value of $L \approx 10$, $k$ is already very close 1 and so $A_{th}$ is very small in magnitude. In fact, it is so small that the variation of the JEF is not sufficient to produce the required phase change from $z = 0$ to $z = L$. The solutions with nodes are not desirable. Just like the case of linear standing waves, the appearance of nodes results in higher energy. In order to find the ground state and at the same time satisfy the boundary conditions, alternative forms of solutions have to be considered. Suppose $A_{th} = i|A_{th}|$, (of course, $A_{th} = -i|A_{th}|$ is also applicable,) then we must have a complex $z_0$ for $\theta$ to be real. In view of the relation $sc(u + i\mathcal{K}') = i/dn(u)$ where $dn(u)$ is another JEF, one must have $\text{Im}(z_0) = \mathcal{K}'/\beta$ where $\mathcal{K}'$ is the elliptic integral of the first kind with modulus $k' = \sqrt{1-k^2}$. Therefore, in the "forbidden region", there are two solutions

$$\theta_a(z) = \frac{1}{\sqrt{K}}\{\frac{\pi}{2} - \varphi + 4\tan^{-1}[i|A_{th}|sc(\beta z - \beta z_a - i\mathcal{K}')]\}$$
$$= \frac{1}{\sqrt{K}}\{\frac{\pi}{2} - \varphi + 4\tan^{-1}[|A_{th}|/dn(\beta z - \beta z_a)]\}, \quad \text{6a}$$

$$\theta_b(z) = \frac{1}{\sqrt{K}}\{\frac{\pi}{2} - \varphi + 4\tan^{-1}[i|A_{th}|sc(\beta z - \beta z_b + \mathcal{K} - i\mathcal{K}')]\}$$
$$= \frac{1}{\sqrt{K}}\{\frac{\pi}{2} - \varphi + 4\tan^{-1}[dn(\beta z - \beta z_b)/|A_{th}|]\} \quad \text{6b}$$

where $z_a$ and $z_b$ are positive constants. Furthermore, we require that $\beta z_b < \mathcal{K}$ to avoid redundancy. That $\theta$ is the phase and $A_{th}$ changes from a real constant to an imaginary constant reminds one of the situation in a lattice. For a periodic potential the eigen states are Bloch functions. As one enters into the energy gap, the states become surface states and the crystal momenta become complex. Here we have similar changes. The solution in eq. (5) can be viewed as the Bloch state because it can be connected to another similar state if the latter is elevated by $\pm 2\pi$ [13]. This kind of connection can be done repeatedly. Hence the solution in eq. (5) is a quasi-periodic function. On the other hand, those in eqs. (6) have a complex argument which corresponding to a complex crystal momentum and above mentioned connection cannot be made. Hence, they can be called edge states in this sense. However, as it will be shown later, there is more subtlety.

For a concrete example, we consider the case of $K = 1/4$ where
case 1, $\tan\alpha_1 \tan\alpha_2 < 0$, or $2n\pi - \pi/2 \leq \varphi \leq 2n\pi + \pi/2$,
case 2, $\tan\alpha_1 \tan\alpha_2 > 0$, or $2n\pi + \pi/2 < \varphi < 2n\pi + 3\pi/2$.

The SGE solutions versus position for different values of $\varphi$ are presented in Figs. 1a and 1b. The solutions in the allowed region are denoted by circle, square and triangle symbols (red) and those of the forbidden region are denoted by diamond, plus and star symbols (green (light) for (6a) and blue (dark) for (6b)) respectively. The black (dashed) lines are the excited states. Their exact forms are

$$\theta(z) = \frac{1}{\sqrt{K}}\{\frac{\pi}{2} - \varphi + 4\tan^{-1}[A_{th}cn(\beta(z-z_0),k)]\}, \qquad 7a$$

and

$$\theta(z) = \frac{1}{\sqrt{K}}\{\frac{\pi}{2} - \varphi + 4\tan^{-1}[1/A_{th}cn(\beta(z-z_0),k)]\} \qquad 7b$$

where $cn(u)$ is again one of the JEFs and $A_{th}^2 = 2k^2/(2k^2 - 1)$ and $\beta^2 = 1/(4k^2 - 1)$. The difference in shape is noticeable. In the "allowed region", the solutions have kinks at both ends and in the forbidden region the solutions exhibit an extremum in the middle of the system. The minimum and maximum occur at $z = z_a$ and $z = z_b$ respectively.

The shape changes of the solutions with $\varphi$ in Fig. 1 are interesting and illuminating. Starting from $\varphi = 0$, the plateau of the solution descends continuously with increasing $\varphi$ until the forbidden region ($\varphi > \pi/2$), is reached. It is a smooth evolution if the solution of eq. (6a) is chosen. On the other hand, beginning at $\varphi = 2\pi$ and reducing $\varphi$, the plateau of the solution rises until $\varphi = 3\pi/2$. At this point the system enters the forbidden region and the solution evolves into the form in eq. (6b). It is remarkable that the system can "chooses" differently when it enters the forbidden region from the different directions.

**4. Energy and level-crossing**

In Fig. 2, we present the energies of the solutions calculated from eq. (2) versus $\varphi$ at $K = 1/4$. The solid (red), dotted (green) and dashed (blue) line correspond to the solutions in eqs. (5), (6a) and (6b) respectively. When $\varphi$ is close to $\pi/2$, the solution in eq. (6a) has the lower energy and when $\varphi$ is close to $3\pi/2$ that of eq. (6b) has lower energy. Indeed we see the system choose the lower energy state in the sense that when the system enters the forbidden region as $\varphi$ increases from $\pi/2$, the solution evolves into that in eq. (6a) and when the system enter the forbidden region as $\varphi$ decreases from $3\pi/2$, the solution evolves into that in eq. (6b).

Notably in Fig. 2, there is a level crossing at a certain value of $\varphi = \varphi_c$. The value of $\varphi_c$ can be evaluated. To have the same energy, it requires that

$$\frac{\partial \theta_a(z)}{\partial z} = \frac{\partial \theta_b(z)}{\partial z}$$

or

$$\frac{\partial \theta_a(z)}{\partial z} = \frac{\partial \theta_b(L-z)}{\partial z}. \qquad 8a$$

For distinct solutions, the later (8a) should been chosen. The boundary conditions in eqs. (4) require that

$$\theta_a(z) = 2\pi - \theta_b(L-z). \qquad 8b$$

Hence, for the r.h.s., we make use of the properties of JEF: $dn(-u) = dn(u)$ and the relation of trigonometric function: $\tan^{-1}x + \tan^{-1}(1/x) = \pi/2$, and reached the interesting relation for the condition of level-crossing:

$$\varphi_c = 3\pi/2 - \sqrt{K}\pi, \qquad 8c$$

and

$$z_a + z_b = L. \qquad 8d$$

In a system with time-reversal symmetry, the level-crossing is the result of Kramers degeneracy. It occurs if the magnetic field vanishes, (in our case $\varphi = \pi/2$). However, in our system there is further complication due to the interaction (appearance of $K$ in eqs. (2) and (3)). In view of eq. (3), it rescales $\theta$ so that we actually have a twisted boundary condition [15]. It is equivalent to inserting a fictitious magnetic flux enclosed by the system. This is exactly the same picture proposed by Laughlin for explaining quantum Hall effect [16]. Hence, it is interesting to note that interaction in our system is equivalent to attaching magnetic flux to particles and as a result, the place of Kramers degeneracy is shifted. This is similar to the situation in ref. 17.

**5. Quantum fluctuation**

Up to now we have only discussed the classical solution of sine-Gordon equation. In this section we would like to assess the effect of the quantum fluctuations on our solutions. We have seen that the solutions in the forbidden region have level crossing. It is possible that the degeneracy is lifted and this phenomenon of classical solitons is altered qualitatively by quantum fluctuations. This problem was first studied by Dashen [18] and Cahill [19] and probably best illustrated by Rajaraman [20].

Our solutions in eqs. (5) and (6) give the potential minima at different values of $\varphi$. For the quantum fluctuations, we consider a small deviation $\eta(z)$ from the solutions in eqs. (6)

$$\tilde{\theta}_{a(b)}(z) = \theta_{a(b)}(z) + \eta_{a(b)}(z), \qquad 9$$

so that $\tilde{\theta}_{a(b)}(z)$ is a quantum solution. By expanding the potential in powers of $\eta_{a(b)}$,

$$V[\tilde{\theta}_{a(b)}] = \int dz [\frac{1}{2}(\partial_z \tilde{\theta}_{a(b)})^2 - \frac{1}{\sqrt{K}}\sin(\sqrt{K}\tilde{\theta}_{a(b)} + \varphi)] \qquad 10$$

we obtained the equation for $\eta_{a(b)}$

$$[-\frac{d^2}{dz^2} + K\sin(\sqrt{K}\theta_{a(b)} + \varphi)]\eta_{a(b)} = w_{a(b)}^2 \eta_{a(b)}. \qquad 11$$

At this stage, it seems that the quantum fluctuations $\eta_a$ and $\eta_b$ are different due to the difference between $\theta_a$ and $\theta_b$. But we can make further transformations. Taking $\theta_a(z)$ as an example, eq. (11) can be written as

$$\{-\frac{d^2}{dz^2} + K[1 - w_a^2 - \frac{8k' dn^2(\frac{z}{1+k'}, k)}{[dn^2(\frac{z}{1+k'}, k) + k']^2}]\}\eta_a = 0. \qquad 12$$

The form can be further simplified.

For transformation mentioned above, consider the phase N=1 solutions of SGE

$$-\theta_{xx} + \sin(\theta) = 0. \qquad 13$$

The meaning of the N phase can be found in ref. 14. It has the form

$$\theta_{N=1}(x) = 2i\ln(f) \qquad 14a$$

where

$$f = \frac{\sqrt{k'_{N=1}}}{dn(\frac{ix}{2\sqrt{k'_{N=1}}}, k_{N=1})}. \qquad 14b$$

Substituting eqs. (14) into eq. (13), one finds

$$-4(\frac{f_x^2}{f^2} - \frac{f_{xx}}{f}) + \frac{1}{f^2} - f^2 = 0, \qquad 15a$$

and integrating eq. (13) one gets
$$f_x^2 = \frac{C}{2}f^2 + \frac{1}{4}(f^4 + 1) \qquad 15b$$
where $C$ is the integration constant. Equations (15) are the useful relations of the JEF $dn$ and will be used to transform eq. (12). However, for $\theta_{N=1}(x)$ and $\theta_a(z)$ (which is a N=2 static solution) to satisfy the same boundary conditions, we have to make the following variable change:
$$\frac{z}{1+k'} = \frac{ix}{2\sqrt{k'_{N=1}}}. \qquad 16$$
The details will be given in appendix. $k$ and $k_{N=1}$ also have a complicated relation between them. As a result, we can write eq. (12) as
$$\frac{4k'_{N=1}}{(1+k')^2} \cdot \frac{d^2\eta_a}{dz^2} + K[1 - w_a^2 - \frac{8}{(f+\frac{1}{f})^2}]\eta_a = 0 \qquad 17$$
with the substitution of eq. (14b). By using the relations of eqs. (14), we get
$$\frac{d^2\eta_a}{dx^2} = \frac{d^2v}{df^2} \cdot (\frac{df}{dx})^2 + \frac{d\eta_a}{df} \cdot \frac{d^2f}{dx^2}$$
$$= \frac{d^2\eta_a}{df^2} \cdot [\frac{C}{2}f^2 + \frac{1}{4}(f^4 + 1)] + \frac{d\eta_a}{df} \cdot [\frac{C}{2}f + \frac{f^3}{2}]. \qquad 18$$
If we choose integral constant $C = 1$, then eq. (17) becomes
$$\frac{4k'_{N=1}}{(1+k')^2}[\frac{(f^2+1)^2}{4} \cdot \frac{d^2\eta_a}{df^2} + \frac{f}{2}(1+f^2) \cdot \frac{d\eta_a}{df}] + K[1 - w_a^2 - \frac{8f^2}{1+f^2}]\eta_a = 0.$$
With the substitution $f = \tan(\phi_a/4)$, above equation becomes
$$\frac{k'_{N=1}}{(1+k')^2} \cdot \frac{d^2\eta_a}{d(\frac{\phi_a}{4})^2} + K[1 - w_a^2 - 8\sin^2(\frac{\phi_a}{4})]\eta_a = 0. \qquad 19$$
Note now that
$$\phi_a = 4\tan^{-1}[\sqrt{k'_{N=1}}/dn(\frac{ix}{2\sqrt{k'_{N=1}}}, k_{N=1})] = 4\tan^{-1}[\sqrt{k'}\, nd(\frac{z}{1+k'}, k)]$$
$$= \varphi - \frac{\pi}{2} + \sqrt{K}\theta_a(z) \qquad 20$$
is just the N=2 static breather solution of SGE!

Exactly the same procedure can be applied to $\theta_b(z)$ with the setting
$$f' = \frac{dn(\frac{ix}{2\sqrt{k'}}, k_{N=1})}{\sqrt{k'_{N=1}}}. \qquad 21$$
As a result we get
$$\frac{k'_{N=1}}{(1+k')^2} \cdot \frac{d^2\eta_b}{d(\frac{\phi_b}{4})^2} + K[1 - w_b^2 - 8\sin^2(\frac{\phi_b}{4})]\eta_b = 0 \qquad 22$$
where $\phi_b = 4\tan^{-1}f'$. Both eqs. (19) and (22) are the well-known Mathieu equation. Its solutions can be calculated with continued fraction [21]. It is interesting to note that their solutions include discrete ones and a continuous spectrum. The discrete ones, called Mathieu functions, are even and odd functions respectively, and are usually denoted as $S_e(\theta)$ and $S_o(\theta)$. Hence, the quantum fluctuations can be evaluated without difficulty.

Equations (19) and (22), both come from eq. (11) but with different classical solutions, have the exactly the same form and hence, the same set of eigen values $w_{a(b)}$. Consequently, we conclude that the quantum fluctuations give the same energy modification to $\theta_a(z)$ and $\theta_b(z)$. At level-crossing, more care should be taken. The zero modes have to be analyzed in order to see if they can lift the degeneracy. In ref. [20] the quantum fluctuation energy modification is written as

$$\Delta E_{QF} = \frac{A}{K} + \frac{1}{2}\sum_{n=1}(\pi_n^2 + q_n^2\omega_n^2) \qquad 23$$

where $\pi_n$ and $q_n$ are the momenta and coordinates operators and $\omega_n$ are the eigen values of eq. (19) and (22). The first term is the contribution of the zero mode and

$$A = \int_0^L (\frac{d\theta_{a(b)}}{dz})^2 dz. \qquad 24$$

At the level-crossing point, in view of eqs. (6,19,22) we found that the contributions of the zero modes for $\theta_a$ and $\theta_b$ are the same. Hence, the degeneracy is not lifted.

Another implication of the quantum fluctuations is that $\tilde{\theta}_a(z)$ and $\tilde{\theta}_b(z)$ do have overlap. The system can switch from one state to the other. This will be important to spin transport which will be discussed later. We close this section by noting that the quantum fluctuation of the solutions in the allowed region eq. (5) can be found with the same procedure.

### 6. Cycles of solitons

In this section we show how the soliton solutions vary adiabatically with $\varphi$. We start for example, from $\varphi = \pi$, the level-crossing point, and take $\theta_b(z)$. As $\varphi$ increases, the soliton solution evolves accordingly, as it can be seen in Fig. 1b. Upon reaching the allowed region boundary at $\varphi = 3\pi/2$, it evolves into $\theta(z)$ in eq. (5). As $\varphi$ increases further, we move across the entire allowed region until we reach $\varphi = 5\pi/2$. For $\varphi > 5\pi/2$, we again enter the forbidden region and the $\theta(z)$ varies smoothly into the ground state $\theta_a(z)$ in eq. (6a). The variation is shown in Fig. 1a. As $\varphi$ increases to $3\pi$, the level-crossing point is reached again. Thus, the evolution of the soliton solution can be divided into three stages:

1. $\pi < \varphi < 3\pi/2$, in the forbidden region, $\theta_b(z)$ in eq. (6b),
2. $3\pi/2 \leq \varphi \leq 5\pi/2$, in the allowed region, $\theta(z)$ in eq. (5),
3. $5\pi/2 < \varphi < 3\pi$, in the forbidden region, $\theta_a(z)$ in eq. (6a).

In view of Fig. 1, it can be concluded that the entire evolution process is smooth. Specifically, at the boundary $\varphi = 3\pi/2$, the solution labeled by diamonds changed into that labeled by triangles as shown in Fig. 1b. It is also true for another boundary at $\varphi = 5\pi/2$ (equivalent to $\pi/2$) where the curve labeled by triangles changes into that labeled by diamonds as shown in Fig. 1a. There are two points worth mentioning. First, in spite of increasing by $2\pi$ in $\varphi$, the soliton solution has not evolved back to its starting form. Second, the energy has been brought back to the initial value.

Upon reaching the level-crossing point $3\pi$, classically nothing happens and hence, the system remains in state described by $\theta_a(z)$ but now it is an excited state. The excited state in the forbidden region evolves into the excited state (see eqs. (7)) in the allowed region at $7\pi/2$. The difference from the ground state is that now $\beta L \approx 2\sqrt{K}\mathcal{K}$. However, it is also possible that due to the quantum fluctuation, the soliton takes a transition from $\theta_a(z)$ to $\theta_b(z)$ at the level-crossing point and thus, the soliton solution goes through a full cycle when $\varphi$ increase by $2\pi$ (from $\pi$ to $3\pi$).

### 7. Spin transport

We prepare the discussion of spin transport with the variation of $\varphi$ by defining spin polarization. We used the definition of spin polarization by Shindou [5]:

$$P_{S^z} = (1/L)\int_0^L z S^z dz \qquad 25$$

This definition is similar to that of charge polarization. It can manifestly shows the spin accumulation in the system of zero magnetization. For example, if $S^z$ is equal to $-1/2$ in the range $0 < z < 1$, $1/2$ in the range $L - 1 < z < L$, and vanishes elsewhere, then $P_{S^z} = 1/2 - 1/4L$. It is easier to see the spin transport by the relation

$$S_j^z = \frac{\sqrt{K}}{2\pi}\frac{\partial\theta}{\partial z} - \frac{(-1)^j\sqrt{K}}{\pi}\sin(\sqrt{K}\theta). \qquad 26$$

Integration by parts gives

$$\delta P_{S^z} = P_{S^z}(\varphi_2) - P_{S^z}(\varphi_1)$$
$$\simeq -\frac{\sqrt{K}}{2\pi L}\{[\int_0^L \theta(z)dz]|_{\varphi_2} - [\int_0^L \theta(z)dz]|_{\varphi_1}\} \qquad 27$$

where the limit of $L \to \infty$ is taken.

From Fig. 1, it is found that in the "allowed region", the plateau of $\theta$ descends or ascends with varying $\varphi$, an indication of the change of spin polarization. By the slopes of the curves, we know that as $\varphi$ increases from 0 (see Fig. 1a), the up-spins at the left end gradually disappear and those at the right end increase, indicating a shift of spin polarization. However, one also sees that the spins hardly change in the inner part (bulk) of the spin chain. Hence, it can be deduced that this spin polarization shift, or spin transport, is due to the edge states.

In the "forbidden region", it is the movement of the extrema of $\theta$ which exhibits spin transport. For example in Fig. 1, as $\varphi$ increases from $3\pi/4$ to $5\pi/4$, the extremum moves. This is another spin polarization shift but involving the bulk spin change. This part of the spin transport is apparently related to the bulk states.

Interestingly, in the "allowed region" the spins at both ends vary (see Figs. 1 and eq. (26)). This suggests that the spins are affected by the environment and the edge states play the role of transporting spins. In the "forbidden region", there is motion of interior spins and the bulk states take up the task. Hence, the meaning of bulk and edge states are contrary to that in a crystal.

One can calculate the change of the spin polarization with eq. (27). Starting from $\varphi = \varphi_c + \delta$, with $\delta$ being an infinitesimal positive number, the system takes the solution of eq. (6b) and evolves with increasing $\varphi$ as described in section 6. When $\varphi = \varphi_c + 2\pi - \delta$, the system is in the state $\theta_a$ and level-crossing is reached. If the system does not make transition from $\theta_a$ to $\theta_b$, then by eqs. (6,8,27)

$$\delta P_{S^z} = -1 + \frac{\sqrt{K}}{\pi L}[\int_0^L \theta_a(z)dz]|_{\varphi_c}. \qquad 28$$

This is not a quantized quantity. However, if there is a transition from $\theta_a$ to $\theta_b$ due to quantum fluctuation, then the spin polarization calculated with eq. (27) gives $\delta P_{S^z} = -1$. Hence, the spin transported in this interval is $\Delta S = 1$ for a full period of $\Delta \varphi = 2\pi$. Importantly, bulks states and edge states combined to give a complete cycle. This quantum spin transport will not be affected by any perturbation, leads or environment. This is also true if we start from any other value of $\varphi$ other than $\varphi_c$.

The origin of the above mentioned quantum spin transport is the topological property of the solitons. Under the adiabatic variation of the external fields, the soliton states change accordingly. When the phase increases by $2\pi$, the soliton states in general do not go back to the original states because they gain energy. However, once the quantum fluctuation-induced transition occurs, the soliton stats indeed will reture to their original states and hence, a quantum spin of unity is transported. Therfore, if one is able to vary the soliton state with a path of closed loop in its parameter space so that it returns to the origial state, quantum spin transport will occur. This is of course related to the winding number in the parameter space.

**8**. **Conclusion**

In conclusion, We have found many solutions of the sine-Gordon equation with finite length. We have also analyzed the quantum spin transport scenario of an antiferromagnetic spin chain. We are able to show that the total spin transported when $\varphi$ increases by $2\pi$ is $\Delta S = 1$ when the quantum fluctuation enables the system makes transition between $\theta_a$ and $\theta_b$. This quantum spin transport is viable in spite of the perturbation of the environment.

This work is supported in part by NSC of Taiwan, ROC under the contract number NSC 98-2923-M-002-002-MY3.

**Appendix**

In this appendix we give the details of the transformation between phase N=1 solutions and phase N=2 static solutions of SGE [14]. Since both are genus=1 solutions, it is reasonable to ask whether they are equal. The answer is yes. The following is the proof. We will use the theta functions to finish our proof because they are closely related to JEF. Some useful relations are listed below:

(1). $\theta_3^2(x,\tau) = \theta_3(2x,2\tau) \cdot \theta_3(0,2\tau) + \theta_2(2x,2\tau) \cdot \theta_2(0,2\tau)$,
(2). $\theta_4^2(x,\tau) = \theta_3(2x,2\tau) \cdot \theta_3(0,2\tau) - \theta_2(2x,2\tau) \cdot \theta_2(0,2\tau)$,
(3). $\theta_2(2x,2\tau) = [\theta_3^2(x,\tau) - \theta_4^2(x,\tau)]/\sqrt{2[\theta_3^2(0,\tau) - \theta_4^2(0,\tau)]}$,
(4). $\theta_4(2x,2\tau) = \theta_3(x,\tau) \cdot \theta_4(x,\tau)/\sqrt{\theta_3(0,\tau) \cdot \theta_4(0,\tau)}$.

**Lemma**

$$\theta_{N=1} = 2i\ln\left[\frac{\theta_4(\frac{ix}{4\sqrt{k'_{N=1}}K_{N=1}},\tau_{N=1})}{\theta_3(\frac{ix}{4\sqrt{k'_{N=1}}K_{N=1}},\tau_{N=1})}\right] = 4\tan^{-1}\left[\frac{\theta_4(\frac{x}{2(1+k'_{N=2})K_{N=2}},\tau_{N=2})}{\theta_3(\frac{x}{2(1+k'_{N=2})K_{N=2}},\tau_{N=2})}\right] = \theta_{N=2}.$$

because both $\theta_{N=1}$ and $\theta_{N=2}$ are genus 1 solutions of $-\theta_{xx} + \sin\theta = 0$. In particular, $\theta_{N=1}$ and $\theta_{N=2}$ are phase $N=1$ and $N=2$ static breathers. The needed transformation is $2\tau_{N=1} = 1 - 1/2\tau_{N=2}$.

**Proof**

1. If one makes a transform, $\tau_{N=1} = (1+\tau_\alpha)/2$, then the corresponding modulus and Jacobi elliptic integral have the relation $\sqrt{k'_{N=1}}K_{N=1} = K_\alpha$

2. By the appendix of ref. 22,

$$\left[\frac{\theta_4(\frac{ix}{4K_\alpha},\frac{1+\tau_\alpha}{2})}{\theta_3(\frac{ix}{4K_\alpha},\frac{1+\tau_\alpha}{2})}\right]^2 = \frac{\theta_3(\frac{ix}{2K_\alpha},1+\tau_\alpha)\theta_3(0,1+\tau_\alpha) - \theta_2(\frac{ix}{2K_\alpha},1+\tau_\alpha)\theta_2(0,1+\tau_\alpha)}{\theta_3(\frac{ix}{2K_\alpha},1+\tau_\alpha)\theta_3(0,1+\tau_\alpha) + \theta_2(\frac{ix}{2K_\alpha},1+\tau_\alpha)\theta_2(0,1+\tau_\alpha)}$$

$$= \frac{\theta_4(\frac{ix}{2K_\alpha},\tau_\alpha)\theta_4(0,\tau_\alpha) - i\theta_2(\frac{ix}{2K_\alpha},\tau_\alpha)\theta_2(0,\tau_\alpha)}{\theta_4(\frac{ix}{2K_\alpha},\tau_\alpha)\theta_4(0,\tau_\alpha) + i\theta_2(\frac{ix}{2K_\alpha},\tau_\alpha)\theta_2(0,\tau_\alpha)}$$

$$= \frac{\theta_2(\frac{x}{2K_\alpha},-\frac{1}{\tau_\alpha})\theta_2(0,-\frac{1}{\tau_\alpha}) - i\theta_4(\frac{x}{2K_\alpha},-\frac{1}{\tau_\alpha})\theta_4(0,-\frac{1}{\tau_\alpha})}{\theta_2(\frac{x}{2K_\alpha},-\frac{1}{\tau_\alpha})\theta_2(0,-\frac{1}{\tau_\alpha}) + i\theta_4(\frac{x}{2K_\alpha},-\frac{1}{\tau_\alpha})\theta_4(0,-\frac{1}{\tau_\alpha})}$$

$$= \frac{[\theta_3(\frac{x}{4K_\alpha},-\frac{1}{2\tau_\alpha}) - i\theta_4(\frac{x}{4K_\alpha},-\frac{1}{2\tau_\alpha})]^2}{[\theta_3(\frac{x}{4K_\alpha},-\frac{1}{2\tau_\alpha}) + i\theta_4(\frac{x}{4K_\alpha},-\frac{1}{2\tau_\alpha})]^2}. \qquad \text{A-1}$$

3. If a further transform $\tau_\alpha = -1/\tau_\beta$ is made, then the corresponding Jacobi elliptic integral has the relation $K_\alpha = K_\beta$.

4. Finally, the transform $\tau_\beta = 2\tau_{N=2}$ gives $K_\beta = \frac{1+k'_{N=2}}{2}K_{N=2}$.

Combining (1), (2), (3), (4), we find

$$\theta_{N=1} = 2i\ln\left[\frac{\theta_4(\frac{ix}{4\sqrt{k'_{N=1}}K_{N=1}},\tau_{N=1})}{\theta_3(\frac{ix}{4\sqrt{k'_{N=1}}K_{N=1}},\tau_{N=1})}\right] = 2i\ln\left[\frac{\theta_4(\frac{ix}{4K_\alpha},\frac{1+\tau_\alpha}{2})}{\theta_3(\frac{ix}{4K_\alpha},\frac{1+\tau_\alpha}{2})}\right]$$

$$= 2i\ln\left[\frac{\theta_3(\frac{x}{4K_\alpha},-\frac{1}{2\tau_\alpha}) - i\theta_4(\frac{x}{4K_\alpha},-\frac{1}{2\tau_\alpha})}{\theta_3(\frac{x}{4K_\alpha},-\frac{1}{2\tau_\alpha}) + i\theta_4(\frac{x}{4K_\alpha},-\frac{1}{2\tau_\alpha})}\right]$$

$$= 2i\ln\left[\frac{\theta_3(\frac{x}{4K_\beta},\frac{\tau_\beta}{2}) - i\theta_4(\frac{x}{4K_\beta},\frac{\tau_\beta}{2})}{\theta_3(\frac{x}{4K_\beta},\frac{\tau_\beta}{2}) + i\theta_4(\frac{x}{4K_\beta},\frac{\tau_\beta}{2})}\right]$$

$$= 2i\ln\left[\frac{\theta_3(\frac{x}{2(1+k'_{N=2})K_{N=2}},\tau_{N=2}) - i\theta_4(\frac{x}{2(1+k'_{N=2})K_{N=2}},\tau_{N=2})}{\theta_3(\frac{x}{2(1+k'_{N=2})K_{N=2}},\tau_{N=2}) + i\theta_4(\frac{x}{2(1+k'_{N=2})K_{N=2}},\tau_{N=2})}\right] = \theta_{N=2} \qquad \text{A-2}$$

We finish the proof by noting that $\tau_{N=1} = (1+\tau_\alpha)/2$, $\tau_\alpha = -1/\tau_\beta$, $\tau_\beta = 2\tau_{N=2}$ and then

$$2\tau_{N=1} = 1 - \frac{1}{2\tau_{N=2}} \qquad \text{A-3}$$

**References**

1. See for example, T. Giamarchi, 'Quantum Physics in One Dimension', Oxford University Press (2004).
2. Fritz Gesztesy, "Soliton Equations and their Algebro-Geometric Solutions V1,

Figure captions:

Fig. 1 The solutions in the allowed region are denoted by circle, square and triangle symbols (red) and those of the forbidden region are denoted by diamond, plus and star symbols (green for (6a) and blue for (6b)) respectively. The dashed lines represent the excited states. Solutions evolve smoothly with varying $\varphi$. (a) The solutions evolve from $\varphi = 0$ to $\varphi = \pi/2$, (eq. (5)), then from $\varphi = \pi/2$ to $\varphi = 3\pi/2$, (eq. (6a)) and finally from $\varphi = 3\pi/2$ to $\varphi = 2\pi$, (eq. (7b)). (b) The solutions evolve from $\varphi = 2\pi$ to $\varphi = 3\pi/2$. (eq. (5)), then from $\varphi = 3\pi/2$ to $\varphi = \pi/2$, (eq. (6b)), and finally from $\varphi = \pi/2$ to $\varphi = 0$, (eq. (7a)).

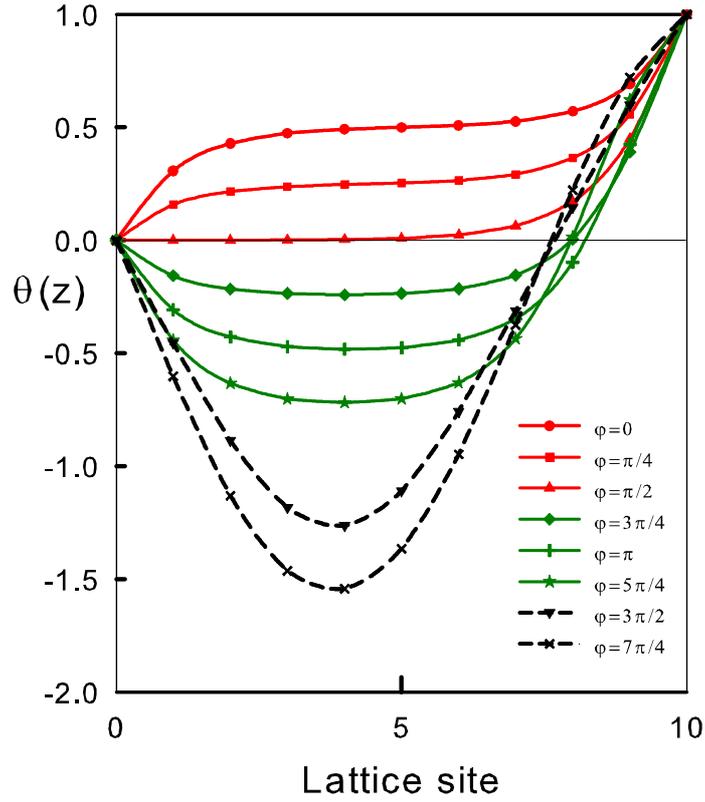

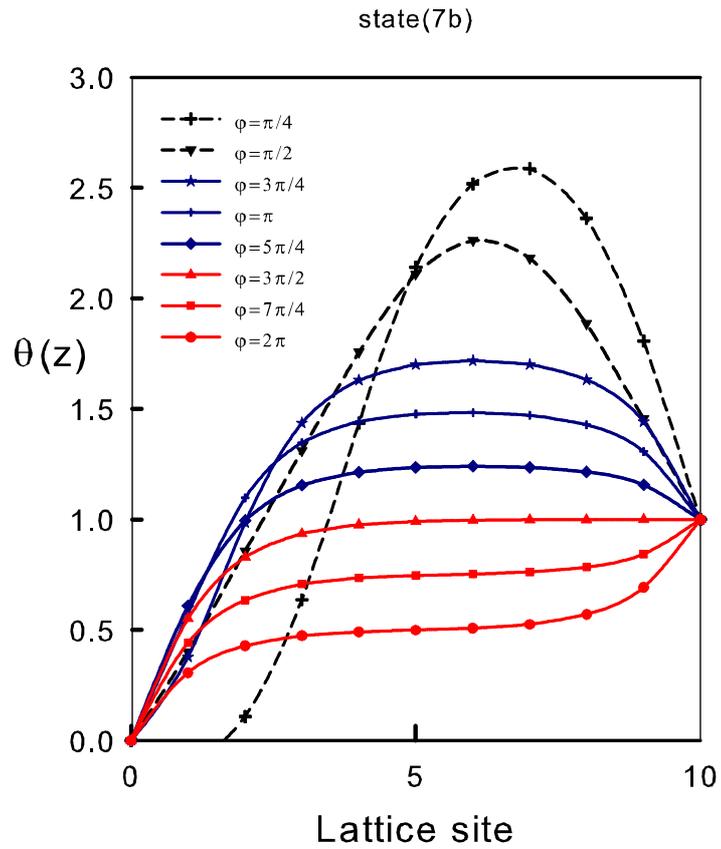

Fig. 2 Energy spectrum (in unit of $\sqrt{vR\sqrt{K}/\pi}$) of the static solitons versus $\varphi$ (in unit of $\pi$)

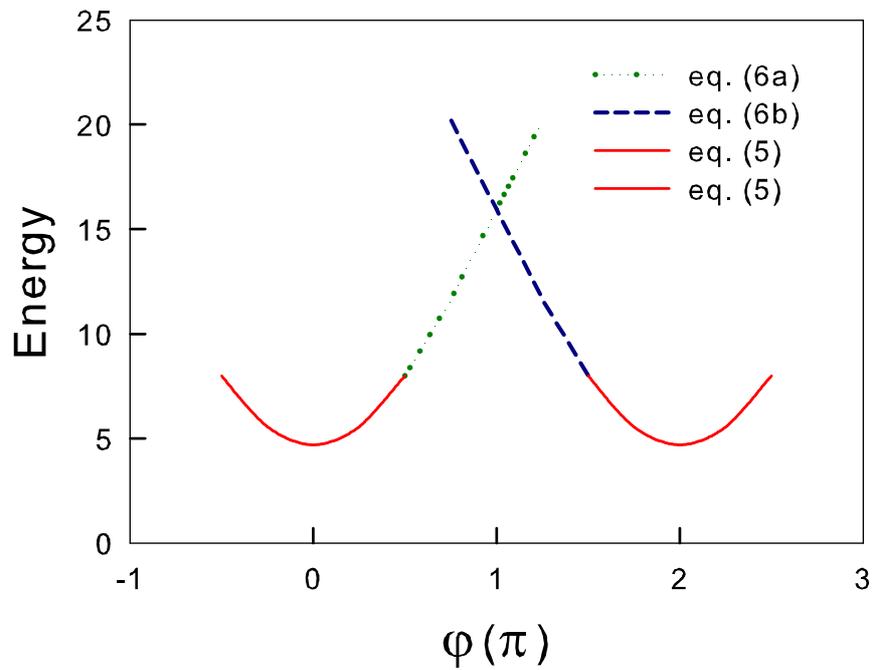